\documentclass[5p,authoryear]{elsarticle}
\usepackage{amssymb,amsmath}
\usepackage{graphicx}
\usepackage{dcolumn}
\usepackage{subfigure}
\usepackage{color}
\usepackage{soul}
\usepackage{float}
\usepackage{times}
\usepackage[breaklinks,colorlinks,
linkcolor=blue,
anchorcolor=blue,
citecolor=blue,
urlcolor=blue,]{hyperref}

\usepackage{color}

\usepackage{xspace}

\renewcommand{\eqref}[1]{\textup{{Eq.~(\ref{#1}})}}
\makeatother

\def\nin{\noindent}

\def\be{\begin{equation}}
\def\ee{\end{equation}}
\def\bea{\begin{eqnarray}}
\def\eea{\end{eqnarray}}

\def\etal{{\it et al.} }

\usepackage{lipsum}
\makeatletter
\def\ps@pprintTitle{%
 \let\@oddhead\@empty
 \let\@evenhead\@empty
 \def\@oddfoot{}%
 \let\@evenfoot\@oddfoot}
\makeatother

\begin{document}

\title{What dominates the time dependence of diffusion transverse to axons: \\Intra- or extra-axonal water?}

\author[nyu]{Hong-Hsi Lee\corref{cor}}
\ead{Hong-Hsi.Lee@nyumc.org}

\author[nyu]{Els Fieremans}
\author[nyu]{Dmitry S. Novikov}

\cortext[cor]{Corresponding author}

\address[nyu]{
Center for Biomedical Imaging, Department of Radiology, NYU School of Medicine, New York, NY 10016}


\begin{abstract}
\nin
Brownian motion of water molecules provides an essential length scale, the diffusion length, commensurate with cell dimensions in biological tissues. Measuring the diffusion coefficient as a function of diffusion time makes in vivo diffusion MRI uniquely sensitive to the cellular features about three orders of magnitude below imaging resolution. However, there is a longstanding debate, regarding which contribution --- intra- or extra-cellular --- is more relevant in the overall time-dependence of the diffusion metrics. Here we resolve this debate in the human brain white matter. 
By varying not just the diffusion time, but also the gradient pulse duration of a standard diffusion pulse sequence, we identify a functional form of the measured time-dependent diffusion coefficient transverse to white matter tracts in 5 healthy volunteers. This specific functional form is shown to originate from the extra-axonal space, and provides  estimates of the fiber packing correlation length for axons in a bundle. Our results offer a metric for the outer axonal diameter, a promising candidate marker for demyelination in neurodegenerative diseases. From the methodological perspective, our analysis demonstrates how competing models, which describe  different physics yet interpolate standard measurements equally well, can be distinguished based on their prediction for an independent ``orthogonal'' measurement.
\end{abstract}

\date{\today}



\maketitle


\section{Introduction}

The ultimate promise of diffusion MRI (dMRI) \citep{jones-book}, a technique that maps the diffusion propagator in each imaging voxel, is to become sensitive and specific to tissue features at the cellular level, orders of magnitude below the nominal imaging resolution. 
The foundation for this sensitivity is provided by the diffusion length, i.e. the rms displacement of water molecules, being of the order of a few $\mu$m,  which is commensurate with cellular dimensions. By controlling the diffusion time, one can probe the time-dependent diffusive dynamics \citep{tanner1979,mitra1992,Assaf200548,RN5,RN4,RN1,RN3,RN2,reynaud2016}, 
and quantify the relevant cellular-level tissue structure indirectly, using biophysical modeling \citep{Yablonskiy_NMRBiomed2010,kiselev2017-review,novikov2016-review}. 


In most tissues, and in the human brain in particular, the dMRI signal generally originates from at least two ``compartments'' --- intra- and extra-cellular spaces \citep{ackerman-neil-nbm2010}. Their distinct microgeometries provide different competing contributions to the overall non-Gaussian diffusion\citep{Assaf200548,RN4,RN5,RN2,RN3}. For any microstructural interpretation of MRI experiments, it is crucial to determine which contribution dominates, and which associated $\mu$m-level length scale can be in principle quantified.


Here we consider diffusion in human white matter (WM), transverse to major WM tracts. For the past decade, the focus of microstructural modeling has been solely on the intra-axonal compartment, where the nontrivial (fully restricted) diffusion was thereby related to the {\it inner} axonal diameters \citep{Assaf200548,RN5,RN4}, whereas the extra-axonal diffusion has been deemed trivial (Gaussian). This framework has served as the basis for a number of techniques (CHARMED \citep{Assaf200548}, AxCaliber \citep{RN5}, ActiveAx \citep{RN4}) for axonal diameter mapping.  
Their outcomes  were subsequently debated due to a notable \citep{RN26}, sometimes by an order-of-magnitude \citep{RN4}, overestimation of human inner axonal diameters relative to their histological values of $\sim 1\,\mu$m \citep{RN9,RN10,RN11,RN13,Tang1997609}. 
This recently prompted an alternative suggestion \citep{RN2,RN3} of the dominant role of non-Gaussian, time-dependent diffusion in the extra-axonal space, with the role of the intra-axonal space deemed trivial (negligible radial signal attenuation due to thin axons). 
Relevant parameters for the extra-axonal picture characterize the packing geometry in a bundle; e.g., the packing correlation length should give a measure of {\it outer} axonal diameters \citep{RN2,RN3}. 

Since both alternatives have compelling arguments behind them and ``fit the data well'' \citep{RN4,RN5,RN2,RN3}, model selection blindly based on  fit quality is unreliable. This is a common challenge of model selection. To address it, here we (i) focus on the {\it functional form} of the competing models originating from their different physical assumptions, and (ii) use the fact that a true model would not just interpolate the standard measurement (varying the diffusion time) where both models perform well, but would also {\it predict} the outcome of an independent ``orthogonal" measurement. For the latter, we vary the gradient pulse width, which was not previously explored. 

Technically, we consider the dependence of the apparent diffusion coefficient $D(\Delta,\delta)$, measured perpendicular to major axonal tracts, both on the diffusion time $\Delta$, and on the diffusion gradient pulse width $\delta$. 
The quantity $D(\Delta,\delta)$ is defined as the lowest-order cumulant term \citep{Kiselev2010diffusion,JensenDKI2005} of the dMRI signal, 
\begin{equation} \label{eq:dmri-signal}
\ln S(\Delta,\delta; g) = - b D(\Delta,\delta) + {\cal O}(b^2)\,, \quad b = g^2\delta^2 (\Delta - \delta/3) \,,
\end{equation}
where $g$ is the applied Larmor frequency gradient, and $b$ is the conventional diffusion weighting \citep{jones-book}.  
The overall $D = f_{\rm in}D_{\rm in} + f_{\rm ex} D_{\rm ex}$ is a weighted average of the apparent intra- and extra-axonal diffusivities, with their $T_2$-weighted fractions normalized to $f_{\rm in} + f_{\rm ex} =1$ 
(we exclude the contribution of myelin water due to its short $T_2\sim 10\,$ms \citep{Mackay1994myelin,RN51d} as compared with our echo time). Remarkably, the functional forms of 
$D_{\rm in}(\Delta,\delta)$ and $D_{\rm ex}(\Delta,\delta)$ will prove to be sufficiently distinct, enabling us to identify which one dominates. 

In the limit $\delta\to 0$, $D(\Delta,\delta)|_{\delta\to0} \to \langle {\bf x}^2(\Delta)\rangle/(2d\Delta)$ corresponds to the genuine water diffusion coefficient in the $d=2$-dimensional plane transverse to the fibers (a weighted average of the genuine compartment diffusivities). 
Finite-$\delta$ measurement imposes a low-pass filter \citep{callaghan,RN3}, suppressing the high-frequency 
dynamics of molecular displacements ${\bf x}(\Delta)$; 
this filter effect is what will technically distinguish $D_{\rm in}(\Delta,\delta)$ and $D_{\rm ex}(\Delta,\delta)$. 
We will use the $\Delta$-dependence to estimate parameters of both models, and then determine which one predicts the ``orthogonal'' $\delta$-dependence best.

\section{Methods}
 
 \subsection{Theory}

We first outline the two models for transverse diffusivity $D(\Delta,\delta)$, paying special attention to their functional forms. 

In the {\it intra-axonal picture}, all $\Delta$- and $\delta$-dependence of 
the radial diffusivity $D\equiv f_{\rm in} D_{\rm in}(\Delta,\delta) + f_{\rm ex} D_\infty^{\rm ex}$ comes from $D_{\rm in}$, 
\begin{equation} \label{eq:D-neu}
D(\Delta,\delta) \simeq D_\infty + \frac{c}{\delta (\Delta-\delta/3)} \,,
\quad 
c = {7\over 48} {f_{\rm in} \bar r^4 \over D_0} 
\end{equation}
based on Neuman's solution \citep{RN7} for narrow impermeable cylinder of radius $r$ (cf. \eqref{eq:D-in-neu} in {\it Appendix A}), with the free (axoplasmic) diffusion coefficient $D_0$; for a distribution of axons, the effective $\bar r^4 \equiv \langle r^6 \rangle/\langle r^2\rangle$ \citep{RN3}. 
Note that \eqref{eq:D-neu} depends on {\it two} independent combinations of tissue parameters: $c$, and 
the overall bulk diffusion coefficient   $D_\infty = f_{\rm ex} D_\infty^{\rm ex}$ (in the  $\Delta\to\infty$ limit); here $D_\infty^{\rm ex}$ is the  bulk diffusion coefficient of the extra-axonal water. 
Typically, $\delta/3 \ll \Delta$; in this limit, the $\sim 1/\Delta$ scaling in \eqref{eq:D-neu} is a consequence of a fully restricted geometry. Less obvious, but crucial for our work, is the inverse scaling with the pulse duration, $D -D_\infty \sim 1/\delta$. It can be traced to the intra-axonal diffusion attenuation $-\ln S_{\rm in} \propto \delta$ inside a cylinder, being equivalent to the effective $T_2^*$ relaxation in the diffusion-narrowing regime \citep{kiselev1998,jensen-chandra-2000,sukstanskii2003,sukstanskii2004,emt-jmr} during the time $\delta$ when diffusion gradients are on; the $1/\delta$ scaling follows from factoring out the $b\propto \delta^2$-dependence, cf. \eqref{eq:dmri-signal}. 

In the {\it extra-axonal picture}, attenuation inside axons is neglected, i.e. $S_{\rm in} \to 1$ and $D_{\rm in} \to 0$, and all dependence of 
$D \equiv f_{\rm ex} D_{\rm ex}$ on $\delta$ and $\Delta$ comes from that of $D_{\rm ex}(\Delta,\delta)$ \citep{RN3,RN2}:
\begin{equation} \label{eq:D-ex}
D(\Delta,\delta) \simeq D_\infty + c' \cdot \frac{\ln (\Delta/\delta) + \frac{3}{2}}{\Delta-\delta/3}\,,
\quad c' = f_{\rm ex} A \,. 
\end{equation}
\eqref{eq:D-ex} is again characterized by two combinations of tissue parameters: $D_\infty$ and $c'$, 
where $D_\infty$ has the same meaning as above, while $c'$ is related to the ``disorder strength'' $A$ characterizing the random packing geometry of axons in the extra-axonal space \citep{RN2,RN3}. 
Here it is crucial that $D$ increases {\it logarithmically} with $1/\delta$, rather than linearly as in \eqref{eq:D-neu}. 
This nontrivial scaling originates from the long-time tail \citep{RN1,Ernst-I,RN3} of the instantaneous diffusion coefficient 
$D_{\rm inst}^{\rm ex}(t) = \frac1{2d} \partial_t \langle {\bf x}^2(t)\rangle \simeq D_\infty^{\rm ex} + A/t$ of the extra-axonal water, restricted by the 
two-dimensional disordered axonal packing geometry; the gradient pulse width $\delta$ provides  short-time cutoff for the tail \citep{RN2,RN3}, which can thereby be probed with varying $\delta$.


\subsection{In vivo MRI}

Diffusion MRI was performed on five healthy subjects (3 males / 2 females, 25-35 years old), by using a 3T Siemens Prisma scanner (Erlangen, Germany) with a 64-channel head coil. The monopolar pulse-gradient spin-echo (PGSE) diffusion tensor imaging (DTI) sequence provided by the vendor (Siemens WIP 511E) was used to perform two different scans for each subject. For each scan, we obtained 3 $b$ = 0 images (no diffusion weighted) and diffusion weighted images (DWI) of $b$ = 0.5 ms/$\mu$m$^2$ along 30 diffusion gradient directions, with an isotropic resolution of (2.7 mm)$^3$ and an FOV of (221 mm)$^2$. The scanned brain volume is a slab of 15 slices, aligned parallel to the anterior commissure (AC) to posterior commissure (PC) line. The corpus callosum was in the middle of the slab, such that the entire corpus callosum was scanned \citep{RN2}. In scan 1, we varied $\Delta$ = [26, 30, 40, 55, 70, 85, 100] ms and fixed $\delta$ at 20 ms; in scan 2, we fixed $\Delta$ at 75 ms and varied $\delta$ = [4, 5, 6.7, 10, 15, 25, 45] ms. All scans were performed with the same TR/TE = 5000/150 ms. Total acquisition time is $\sim$ 50 min.

\subsection{Image processing}

Our image processing pipeline includes four steps: denoising, Gibbs ringing elimination, eddy-current and motion correction, and diffusion tensor estimation.

For denoising, we identified and truncated noise-only principle components by using the fact that principle component analysis eigenvalues, arising from noise, obey the universal \\Marchenko-Pastur distribution \citep{RN42,Veraart2016394}. To eliminate Gibbs ringing, we re-interpolated each denoised image by sampling the ringing pattern at the zero-crossings of the sinc function \citep{RN43}. Then we used FSL eddy to correct eddy-current distortions and subject motions \citep{RN44}.
Finally, diffusion tensors were evaluated via an unconstrained weighted linear least squares (WLLS) method, where the weights were estimated from diffusion tensor calculations based on an unweighted LLS method \citep{RN45}. The contribution of imaging gradients to b-value is negligible since it is always less than $10^{-3}$ ms/$\mu$m$^2$ in our experiments.

If the diffusion data have SNR$\, > 2$, tensor estimations of WLLS will not be biased by Rician noise \citep{RN45}. To calculate the SNR of $b$ = 0 images, denoised signal was divided by the estimated noise level of the noise map, obtained from denoising method mentioned above \citep{RN42}. In our $b$ = 0 images, mean SNR of the WM was $\approx$ 18-22. Considering that WM's $D_\parallel\sim$ 1.2-1.6 $\mu$m$^2$/ms, $D\sim$ 0.5 $\mu$m$^2$/ms and $b$ = 0.5 ms/$\mu$m$^2$, SNR in DWIs was still much higher than 2, and thus WLLS gave us unbiased tensor estimations.

For each voxel, we calculated eigenvalues of the diffusion tensor estimated via WLLS, sorted in the order  $\lambda_1 \geq \lambda_2 \geq \lambda_3$. Axial diffusivity, defined by $D_\parallel \equiv \lambda_1$, estimates diffusion parallel to axons. Similarly, radial diffusivity, defined by $D \equiv (\lambda_2 + \lambda_3)/2$, estimates diffusion transverse to axons. In this way, we obtain  maps of axial diffusivity, radial diffusivity, 
and fractional anisotropy (FA) \citep{Basser1994dti}.

\subsection{Region of Interest (ROI)}

To automatically delineate WM ROIs, we registered each subject's mean FA map to FSL's standard FA map in MNI 152 space with FMRIB's linear image registration tool (FLIRT) and non-linear registration tool (FNIRT) \citep{RN47,RN48,RN49}. The individual mean FA map is acquired by averaging all the FA maps in different $\Delta$ and $\delta$ in scans 1 and 2 for each subject. The transformation matrix (FLIRT) and the warp (FNIRT) were retrieved to inversely transform the WM atlas ROIs from MNI 152 space to the individual subject space. In our study, we used the Johns Hopkins University DTI-based WM atlas \citep{RN50}, which was registered to MNI 152 space with FLIRT and FNIRT before use. To suppress the cerebrospinal fluid (CSF) signal contamination due to the long TE, we used an extended CSF mask to exclude WM voxels close to CSF. The CSF mask was segmented from a mean $b$ = 0 image by FMRIB's Automated Segmentation Tool (FAST) \citep{RN46}, and its edge was expanded by one voxel. One subject's WM ROIs are shown in Fig.~\ref{fig:ex-neu}c. In the scanned slab, we focused on the main WM tracts including anterior corona radiata (ACR), superior corona radiata (SCR), posterior corona radiata (PCR), posterior limb of the internal capsule (PLIC), genu, midbody, and splenium of the corpus callosum. 

\subsection{Data Analysis}

Eigenvalues, axial and radial diffusivities were calculated voxel by voxel and averaged over each ROI. To evaluate the strength of the $\Delta$-dependence described by intra- and extra-axonal models, we assumed that the $D$ in scan 1 is a linear function of $1/(\delta(\Delta-\delta/3))$  and $(\ln(\Delta/\delta)+3/2)/(\Delta-\delta/3)$, suggested by \eqref{eq:D-neu} and \eqref{eq:D-ex}, and calculated the two models' Pearson's linear correlation coefficients $R$ and $P$-values with the null hypothesis of no correlation. If $P  < 0.05$ in an ROI, the null hypothesis is rejected, and the $\Delta$-dependence is non-trivial. In the ROIs with significant $\Delta$-dependence, we fit \eqref{eq:D-neu} and \eqref{eq:D-ex} to the scan 1 data and acquired parameters shown in Table~\ref{tab:ex-neu}.

\begin{table*}[th!!]
{\centering
\begin{tabular}{l | cccccc | ccccc}
\cline{1-12}
& \multicolumn{6}{c|}{Intra-axonal model, \eqref{eq:D-neu}} & \multicolumn{5}{c}{Extra-axonal model, \eqref{eq:D-ex}}\\
\cline{2-12}
ROI & $P$ & $R^2$ & $D_\infty$ & $c$ & $2\bar r\left(\frac{f_{\rm in}}{D_0}\right)^{1/4}$ & $\bar \eta\left(\frac{f_{\rm in}}{D_0}\right)^{1/4}$ & $P$ & $R^2$ & $D_\infty$ & $c'$ & $l_c^\perp \sqrt{f_{\rm ex}}$\\
\cline{1-12}
ACR 	&	2.1e-3	&	0.871	&	0.603	& 43.3 &	5.13		&	1.73		&	1.5e-3	&	0.887	&	0.597	&	0.241	&	1.10\\
SCR 	&	2.5e-4	&	0.945	&	0.523	& 62.3 &	5.62		&	1.90		&	1.7e-4	&	0.952	&	0.515	&	0.338	&	1.30\\
PCR 	&	6.5e-4	&	0.919	&	0.592	& 85.4 &	6.08		&	2.05		&	2.9e-4	&	0.942	&	0.581	&	0.484	&	1.56\\
PLIC 	&	7.0e-4	&	0.917	&	0.427	& 80.8 &	6.00		&	2.03		&	5.9e-4	&	0.922	&	0.419	&	0.427	&	1.46\\
Genu 	&	0.60		&	-		&	-		& -	&	-		&	-		&	0.65		&	-		&	-		&	-		&	-\\
Midbody 	&	0.24		&	-		&	-		& -	&	-		&	-		&	0.31		&	-		&	-		&	-		&	-\\
Splenium 	&	1.2e-3	&	0.896	&	0.349	& 106.7 &	6.43		&	2.17		&	2.1e-3	&	0.873	&	0.337	&	0.560	&	1.67\\
\cline{1-12}
\end{tabular}\\}
\caption{
\label{tab:ex-neu}
Estimated parameters from scan 1, based on intra-axonal model, \eqref{eq:D-neu}, and extra-axonal model, \eqref{eq:D-ex}.
{\it Intra-axonal model:}   
Values of $2\bar r \left(f_{\rm in}/D_0\right)^{1/4}$ and $\bar \eta \left(f_{\rm in}/D_0\right)^{1/4}$ are lower bounds of, respectively, the (volume-weighted) inner axonal diameter $2\bar r$ (cf. text below \eqref{eq:D-neu}), and of the axonal shrinkage $\bar \eta$ (Fig.~\ref{fig:histogram} and \eqref{eq:shrink-factor}) since, practically, $f_{\rm in}/D_0 < 1\,\mathrm{ms/\mu m^2}$. 
{\it Extra-axonal model:} We used empirical estimate \citep{RN3} $A\approx 0.2\, (l_c^\perp)^2$, to obtain 
the combination $l_c^\perp \sqrt{f_{\rm ex}}$ from $c'$. 
This sets a lower bound \citep{RN2} on the fiber packing correlation length $l_c^\perp$ because $f_{\rm ex} < 1$; 
 $l_c^\perp$ provides an estimate for the outer axonal diameter. 
All parameters are in the corresponding units of $\mu$m and ms.
}
\end{table*}

\section{Results}

\begin{figure}[t!]
\centering
\includegraphics[width=.99\linewidth]{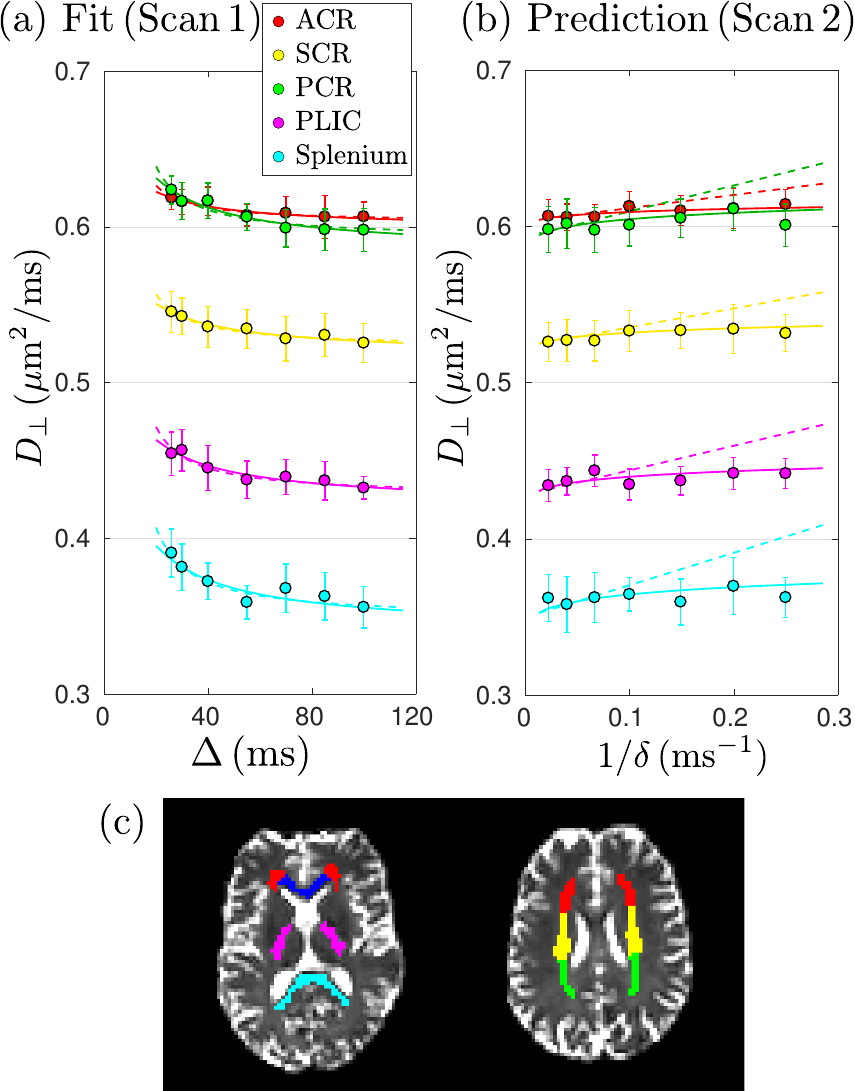}
\caption{Radial diffusivity $D(\Delta,\delta)$ for WM ROIs averaged over five subjects. (a) With fixed $\delta$ = 20 ms, $D$ from scan 1 decreases with $\Delta$. Dashed  and solid lines are fits based on \eqref{eq:D-neu} (intra-axonal) and \eqref{eq:D-ex} (extra-axonal), correspondingly. 
(b) With fixed $\Delta$ = 75 ms, $D$ from scan 2 increases as a function of $1/\delta$. Dashed and solid lines are {\it predictions} (not fits) based on parameters obtained from scan 1 (Table~\ref{tab:ex-neu}), using the corresponding models, \eqref{eq:D-neu} and \eqref{eq:D-ex}, where now $\Delta$ is fixed and $\delta$ varies. 
(c) WM ROIs, including ACR (red) = anterior corona radiata, SCR (yellow) = superior corona radiata, PCR (green) = posterior corona radiata, PLIC (magenta) = posterior limb of the internal capsule, genu (blue), and splenium (cyan) of the corpus callosum.}
\label{fig:ex-neu}
\end{figure}

In Fig.~\ref{fig:ex-neu}, we show the results for brain scans of five healthy subjects with a monopolar PGSE DTI sequence. The mean values of $D$ were computed within each ROI in brain WM,  Fig.~\ref{fig:ex-neu}c, and averaged over five subjects. 

To explicitly reveal the dependence of $D$ on both $\Delta$ and $\delta$, we performed 2 scans for each subject. In {\it scan 1}, we fixed $\delta$ = 20 ms, as it is typically done \citep{RN2,RN18,Barazany2009,Nilsson2009176,Horsfield1994Dt,Stanisz1997Dt,Barshir2008-highb,Kunz2013-uhf}, 
and varied $\Delta$. Scan 1 embodies a standard $t$-dependent ($t\approx \Delta$) dMRI measurement $D(t)$. In {\it scan 2}, we fixed $\Delta$ = 75 ms and varied $\delta$ instead. This $\delta$-dependence has not been comprehensively studied, and turns out to be quite revealing. 

Fig.~\ref{fig:ex-neu}a shows that both the intra- and extra-axonal models fit the ``standard'' scan 1 data well in each ROI. The estimated $P$-value, $R^2$, and fit parameters are shown in Table~\ref{tab:ex-neu}. A naive way to select between the two models would be to use the $R^2$ goodness-of-fit parameter (since both models have the same number of 2 degrees of freedom). However, while $R^2$ is generally closer to $1$ for the extra-axonal model, we feel it is not enough to use this noisy metric to unequivocally select \eqref{eq:D-ex}. For a physically more informed model selection, we now focus on the functional form of the $\delta$-dependence, by using fit parameters ($D_\infty$ and $c$, and $D_\infty$ and $c'$, correspondingly, Table~\ref{tab:ex-neu}),  to predict scan 2 data. Fig.~\ref{fig:ex-neu}b shows that the parameter-free predictions of the two models are very different, both quantitatively and qualitatively; the diffusivity for extra-axonal model, \eqref{eq:D-ex}, captures the systematic bend in the curves with respect to $1/\delta$ very well, while \eqref{eq:D-neu} for intra-axonal model increases linearly with $1/\delta$ and clearly deviates from experimental results. 

We emphasize that the prediction of scan 2 was performed without any adjustable parameters, since tissue properties are found in scan 1, and the $\delta$-dependence is calculated based on \eqref{eq:D-neu} and \eqref{eq:D-ex}. Hence, this prediction provides a parameter-free test of the models involved. 


Fig.~\ref{fig:ex-neu} shows that the extra-axonal model demonstrates better consistency between scans 1 and 2, indicating that the contribution of extra-axonal water dominates the signal change. We can also observe this by inspecting model parameter values. Using fit parameters based on the intra-axonal model (see Table~\ref{tab:ex-neu}, \eqref{eq:D-neu}, and {\it Appendix A, Wide pulse limit in the GPA}) and typical values of $f_{\rm in} \approx 0.5$ and  $D_0 \gtrsim  1.5\, \mu$m$^2$/ms \citep{rotinv}, the estimated inner axonal diameter $2\bar r \approx 6.8-8.5\,\mu$m, much larger than histologically reported  values $\approx 1\,\mu$m  \citep{RN9,RN10,RN11,RN13,Tang1997609}.

Based on fit results of extra-axonal model (see Table~\ref{tab:ex-neu}, \eqref{eq:D-ex}, and $A\approx 0.2\,(l_c^\perp)^2$ from ref.~\citep{RN3}), and  $f_{\rm ex} \approx 0.5$, we estimate the axonal packing correlation length $l_c^\perp\approx1.6-2.4\,\mu$m. As typical values of the ratio of inner to outer diameter (the g-ratio) range within $0.6-0.8$  in central nervous system \citep{RN12,Stikov2015-gratio}, the outer axonal diameter $\sim 1\,\mu$m/(g-ratio) $\approx 1.3-1.7\,\mu$m, close to estimates of the correlation length in our experiments. Tang and Nyengaard \citep{RN13,Tang1997609} uniformly sampled the WM of one human brain hemisphere and also reported the outer diameter of myelinated axons of about 1.14 $\mu$m on average. The scale of the fiber packing correlation length is biologically plausible, and could be a potential biomarker for the outer diameter, a metric of myelination, which is an important hallmark of neurodegeneration, such as multiple sclerosis \citep{Bando201516}.

\section{Discussion}

By varying both $\Delta$ and $\delta$, and identifying physical origins of these dependencies, our {\it in vivo} dMRI measurements distinguish between functional forms of intra- and extra-axonal models, and show the predominance of the extra-axonal time dependence in human brain WM.  The extra-axonal model offers an estimate of outer axonal diameter via packing correlation length, whose changes can be sensitive to demyelination, and possibly axonal loss or other kinds of geometric changes in axonal fiber tracts at the $\mu$m level, three orders of magnitude below the achievable resolution of human MRI.

In what follows, we will put our work in context of previous measurements using shorter times or thicker axons (in the spinal cord), and employing larger gradients, as well as discuss a possible relation between the disorder strength $A$ characterizing outer axonal diameters, and the measurements of axonal conduction velocity.

\begin{figure}[b!!]
\centering
\includegraphics[width=.99\linewidth]{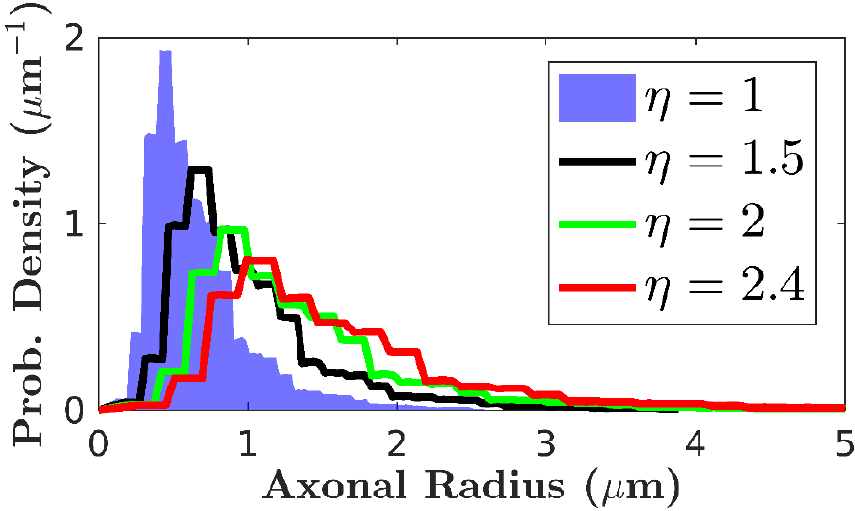}
\caption{Histogram of axonal radii $h_i = h(r_i)$, based on histological results in corpus callosum of three post-mortem human brains \citep{RN10} sampled into 100 bins $r_i$. The shrinkage factor $\eta$  extends the bins $r_i \to \eta r_i$, modeling a correction for the axonal radii due to a uniform tissue shrinkage during fixation, with $\eta$ = 1 (blue area) corresponding to no shrinkage, i.e. the measured histogram equal to that in vivo.}
\label{fig:histogram}
\end{figure}

\subsection{Intra-axonal model: when pulses are not wide}

\begin{figure*}[th!!]
\centering
\includegraphics[width=.8\linewidth]{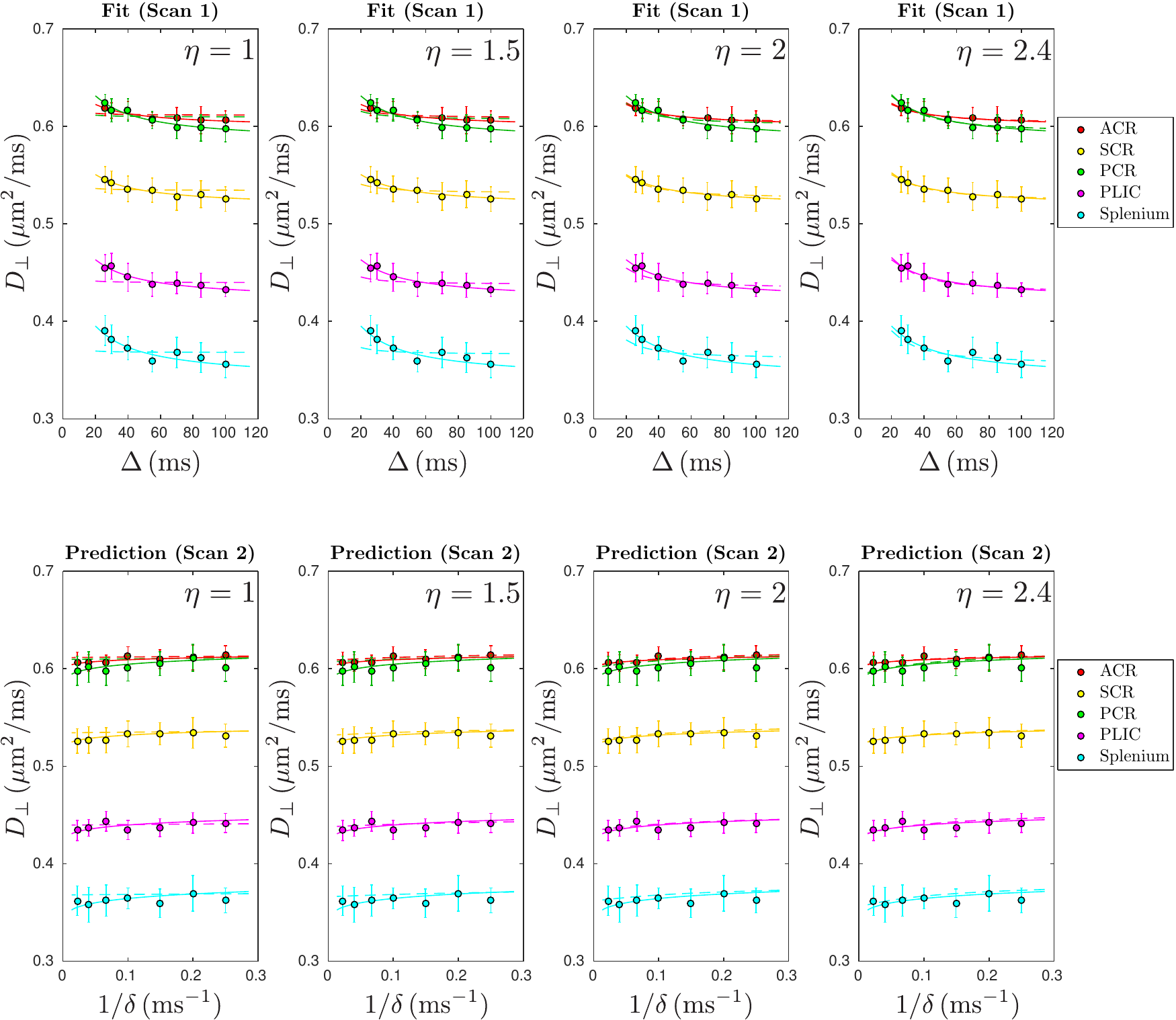}
\caption{$D$ from Fig.~\ref{fig:ex-neu}, fit with full intra-axonal model (van Gelderen), \eqref{eq:S-in-vgel} and \eqref{eq:D-vgel}, for the four values of the shrinkage factor $\eta$ = [1, 1.5, 2, 2.4] (dashed lines, top row). 
Poor fits for $\eta\leq 2$ are due to $f_{\rm in}$ hitting the upper bound. 
Dashed lines in bottom row are predictions (not fits) for $\delta$-dependence in scan 2 data, based on parameters obtained from scan 1 (Table~\ref{tab:vgel} and Eq.~\ref{eq:D-vgel}). 
Solid lines in upper and lower rows are the same as those in Fig.~\ref{fig:ex-neu}, i.e. fits for scan 1 and predictions for scan 2 based on Eq.~\ref{eq:D-ex} (extra-axonal model), shown here for reference.
}
\label{fig:vgel-wm}
\end{figure*}

Suppose, for a moment, that despite all the above arguments, the intra-axonal model is the true one. Then, according to our results in Table~\ref{tab:ex-neu}, the very large inner axonal radius $\bar r \approx 4 \mu$m should lead to an intra-axonal correlation time (time to diffuse across an axon) $t_{\rm c} = r^2 / D_0 \approx\, 10\,$ms. Technically, \eqref{eq:D-neu} is applicable only if the wide pulse limit ($\delta \gg t_c$) is satisfied (see details in {\it Appendix A}). For histologically feasible $2\bar r \sim 1\,\mu$m, \eqref{eq:D-neu} applies, since $t_c < 1\,$ms, and 
$\delta$ in scan 2 varied from 4-45 ms; that is the reason we used the Neuman's approximation in \eqref{eq:D-neu}. 
However, for the ``apparent'' $t_c$ based on fits of \eqref{eq:D-neu} to scan 1 data,  the wide pulse limit is violated. 
Hence, we will repeat our intra-axonal model analysis using  a more general, albeit less analytically transparent equation due to van Gelderen \etal \citep{RN6}, applicable to axons of all sizes, and will employ the axonal radius histogram, Fig.~\ref{fig:histogram}, according to histological observations \citep{RN10} (cf. \eqref{eq:S-in-vgel} and \eqref{eq:D-vgel} in {\it Appendix A}). 
We will refer to this modified model as the {\it intra-axonal model (van Gelderen)}.

Very large apparent axonal diameters would necessarily imply strong brain tissue shrinkage in fixation and paraffin embedding \citep{horowitz2015response}, such that  histologically measured axons have to be assumed notably smaller than in vivo.   
To compensate for such hypothetical shrinkage, we introduce a shrinkage factor $\eta > 1$, which linearly extends the measured radii histogram (Fig.~\ref{fig:histogram} and {\it Appendix A}), such that mean axonal radius is $\eta$ times larger than that calculated with histology. We note from the outset, that $\eta$ cannot exceed $1.5$, as argued in refs.~\citep{RN9,RN15}, and $\eta \sim 1.03 - 1.07$ measured in  ref.~\citep{RN13,Tang1997609}.

Fig.~\ref{fig:vgel-wm} shows that, in each ROI, the full intra-axonal model (van Gelderen), \eqref{eq:S-in-vgel} and \eqref{eq:D-vgel}, can neither fit the scan 1 data nor predict the $\delta$-dependence in scan 2 data if $\eta\leq 2$. 
(We fixed $\eta$ to a few values, instead of letting it vary, to achieve fit robustness.) 
 Based on the parameters in Table~\ref{tab:vgel}, in most of the ROIs, the values of $f_{\rm in}$ hit the upper bound and the fits are poor ($R^2 <$ 0.9) if $\eta\leq$ 2. Thus, to fit the data with reasonable parameters and to predict the $\delta$-dependence, the shrinkage due to tissue fixation should exceed two-fold, which contradicts available histological data \citep{RN9,RN15,RN13,Tang1997609}. 
 
Interestingly, when $\eta >2$, the functional form of the intra-axonal model (van Gelderen) is very similar to that of our extra-axonal model, i.e. the intra-axonal model begins to describe both the varying $\Delta$ and $\delta$ data sets equally well. This explains why, in previous studies, which were performed at a few $\Delta$ and $\delta$ and which did not focus on the functional form of $D(\Delta,\delta)$, the axonal diameter estimations based on the intra-axonal model alone were much larger than that in histological studies \citep{RN4,Barazany2009} --- the fitting was ``stretching'' the axons to match the data. In contrast, the extra-axonal model, \eqref{eq:D-ex}, does not stretch the length scales, and provides precise predictions for the $\delta$-dependence, Fig.~\ref{fig:ex-neu}b, and realistic packing correlation length estimates, Table~\ref{tab:ex-neu}. 

We also note that for the spinal cord, where axons are about factor of 5 thicker than those in the brain, one must use the full van Gelderen's model since $t_c \sim 10\,$ms. In this situation, the balance between the intra- and extra-axonal time-dependencies should be revisited, due to the very strong, $\sim r^4$ scaling of the intra-axonal signal, so that both effects are now comparable. The dMRI measurement can become sensitive to the inner diameters of the spinal cord WM, and reasonable diameter estimates can be obtained \citep{Benjamini2016333,Komlosh2013210};  
however, accounting for the nontrivial time-dependence of the extra-axonal diffusion coefficient still improves such estimates \citep{xu2014}.

\begin{table*}[th!]
{\centering
\begin{tabular}{l | ccc | ccc | ccc | ccc}
\cline{1-13}
& \multicolumn{12}{c}{Intra-axonal model (van Gelderen), \eqref{eq:D-vgel}}\\
\cline{2-13}
ROI	& \multicolumn{3}{c |}{$\eta=1$} & \multicolumn{3}{c |}{$\eta=1.5$} & \multicolumn{3}{c |}{$\eta=2$}  & \multicolumn{3}{c}{$\eta=2.4$}\\
\cline{2-13}
	& $R^2$ & $D_\infty$ & $f_{\rm in}$ & $R^2$ & $D_\infty$ & $f_{\rm in}$ & $R^2$ & $D_\infty$ & $f_{\rm in}$ & $R^2$ & $D_\infty$ & $f_{\rm in}$\\
\cline{1-13}
ACR		&	0.145	&	0.611	&1*	&	0.562	&	0.608	&1*	&	0.872	&	0.603	&0.901	&	0.875	&	0.602	&0.510\\
SCR		&	0.112	&	0.534	&1*	&	0.463	&	0.531	&1*	&	0.903	&	0.525	&1*		&	0.950	&	0.522	&0.723\\
PCR		&	0.076	&	0.610	&1*	&	0.327	&	0.607	&1*	&	0.733	&	0.600	&1*		&	0.933	&	0.592	&1*\\
PLIC		&	0.071	&	0.440	&1*	&	0.312	&	0.438	&1*	&	0.719	&	0.433	&1*		&	0.931	&	0.427	&1*\\
Splenium	&	0.047	&	0.368	&1*	&	0.211	&	0.366	&1*	&	0.527	&	0.361	&1*		&	0.785	&	0.354	&1*\\
\cline{1-13}
\multicolumn{13}{l}{* The fitting parameter $f_{\rm in}$ hits the upper bound.}
\end{tabular}\\}
\caption{
\label{tab:vgel}
Estimated parameters from scan 1, based on intra-axonal model (van Gelderen), \eqref{eq:D-vgel}, fixed at four shrinkage factors $\eta$. The range of $f_{\rm in}$ is [0, 1]. In most of the ROIs, when $\eta \leq 2$, the fitted $f_{\rm in}$ hits its upper bound, and the fit is poor ($R^2 < 0.9$), which is also shown in the upper rows of Fig.~\ref{fig:vgel-wm}. To obtain a better fit and ensure $f_{\rm in} < 1$, shrinkage factor $\eta$ needs to exceed 2, which is unrealistic \citep{RN9,RN10,RN11,RN15,RN13,Tang1997609}. 
}
\end{table*}

\subsection{Relation to measurements with strong diffusion gradients}

Applying extremely strong diffusion gradients $G\sim 0.1-1\,$T/m facilitates the estimation of intra-axonal parameters \citep{Barazany2009,RN17,RN18,RN4,RN5,RN16}
because of stronger signal attenuation inside axons, as well as due to exponential suppression of the extra-axonal signal in the radial direction, roughly as $\sim f_{\rm ex}\,e^{-bD_\infty^{\rm ex}}$. 

However, for strong diffusion gradients 
the intra-axonal model needs corrections, since the Gaussian phase approximation (GPA) for $S_{\rm in}$, under which both Neuman's and van Gelderen's solutions were obtained, eventually breaks down. Unfortunately, no solutions beyond GPA currently exist for finite pulse width $\delta$. 
In {\it Appendix A, Beyond GPA}, we estimate that 
GPA breaks down when 
\begin{equation} \label{eq:gpa-in-condition}
g \gtrsim g^* = \frac{D_0}{r^3} = \frac{1}{r} \cdot \frac{1}{t_{\rm c}}\,.
\end{equation}
The Larmor frequency gradient $g\equiv \gamma G$ is defined via the proton gyromagnetic ratio $\gamma$. 
For reference, $g= 0.0107\, \mathrm{(\mu m\cdot ms)^{-1}}$ for $G = 40\,$mT/m (typical human scanner). 


Estimating Larmor frequency inhomogeneity across an axon by $\Omega \sim g^* \cdot r$, the above condition becomes $\Omega \cdot t_{\rm c} \sim 1$, i.e. the typical precession phase during diffusion across an axon is $\sim 1$ (i.e. not small).  Note that the critical gradient $g^*$ is purely determined by tissue properties, independent of sequence timings. For example, if $r = 3\,\mu$m and $D_0 = 1.5\,\mu$m$^2$/ms, $g^* = 0.0556\, \mathrm{(\mu m\cdot ms)^{-1}}$ (corresponding to $G = 208\,$mT/m); when the actual $g$ becomes of this order of magnitude (and proportionally larger for smaller axons), the higher-order in $g$ corrections to GPA become crucial. 

In our experiments, the gradient strength stays below 77\,mT/m, and  GPA perfectly applies. Recent studies boosted diffusion gradients up to $G\lesssim 300\,$mT/m for humans \citep{RN16} and $G\lesssim 1.3\,$T/m for ex-vivo mice \citep{RN17}. When the signal contribution of large axons is not negligible, beyond-GPA corrections are needed due to the tail of axonal histogram extending to large axons, 
since for them, the critical $g^*$ decreases as $1/r$. 
The {\it negative} beyond-GPA correction to $\ln S_{\rm in}$, \eqref{eq:gpa-in-phi}, may therefore explain the residual overestimation of axonal diameters in the study \citep{RN17} with ultra-strong gradients --- basically, this correction tells that $S_{\rm in}$ experiences extra attenuation due to the ${\cal O}(g^4)$ contribution, neglected in standard axonal diameter mapping frameworks.

Similarly, higher-order corrections in the powers of diffusion weighting $b\propto g^2$, \eqref{eq:dmri-signal}, 
should be considered for the extra-axonal signal. 
The extra-axonal signal $S_{\rm ex}$ up to ${\cal O}(b^2)$ can be obtained from the recent narrow-pulse result [Appendix E of ref.~\citep{RN3}], by substituting $t_c\to \delta$ as the logarithmic cutoff:  
\begin{equation} \label{eq:S-ex-kurtosis}
\ln S_{\rm ex} \simeq 
- b \, D_{\rm ex}(\Delta,\delta) 
+ \frac{K_{\rm ex}}{6} (b D_\infty^{\rm ex})^2\,,
\end{equation}
where 
$K_{\rm ex}(\Delta,\delta)$ is the apparent extra-axonal kurtosis,
\begin{equation} \notag
\frac{K_{\rm ex}(\Delta,\delta)}{6} \simeq \frac{A}{D_\infty^{\rm ex}} \cdot \frac{\ln(\Delta/\delta)}{\Delta}\,, \quad \Delta\gg \delta \gg t_c \,. 
\end{equation}
Here, the genuine kurtosis $K_{\rm ex}(t)$ has the $\ln(t/t_c)/t$ tail \citep{RN3}, 
and we used the low-pass filter analogy in the wide pulse limit $\delta\gg t_{\rm c}$,
to re-define the long-time tail cut-off, $t_c \to \delta$.
 
\eqref{eq:S-ex-kurtosis} tells that the ${\cal O}(b^2)$ kurtosis term becomes of the order of the nontrivial, time-dependent ${\cal O}(b)$ term, when 
$bD_\infty^{\rm ex} \gtrsim 1$; this condition practically coincides with the breakdown of the ${\cal O}(b)$, DTI representation 
$\ln S \approx -b\cdot f_{\rm ex}D_\infty^{\rm ex}$, for the total signal 
$S \simeq f_{\rm in}  + f_{\rm ex} \, e^{ - bD_\infty^{\rm ex}}$.    
In other words, at the same $b$ when the curvature of the observed $\ln S$ versus $b$ becomes notable, the extra-axonal $K_{\rm ex}$ term in \eqref{eq:S-ex-kurtosis} must be included in the analysis if one wants to estimate $A$ and $f_{\rm in}$, $f_{\rm ex}$ (and, possibly the inner radii)  separately, by going to high $b$; one {cannot} use the approximation 
$S|_{K_{\rm ex}\equiv 0} \simeq f_{\rm in} S_{\rm in} + f_{\rm ex}\, e^{-bD_{\rm ex}(\Delta,\delta)}$  
beyond its ${\cal O}(b)$ term. 
De Santis \etal \citep{RN18} used the $S|_{K_{\rm ex}\equiv 0}$ approximation to modify AxCaliber estimation of inner diameters from human brain data in the corpus callosum acquired with stimulated echo dMRI with $b\leq 4\,\mathrm{ms/\mu m^2}$.  
Including the $D_{\rm ex}(\Delta,\delta)$ term in \eqref{eq:S-ex-kurtosis} resulted in about 5-fold smaller inner diameter estimates in comparison to just using $D_\infty^{\rm ex}$, effectively demonstrating the importance of non-Gaussian (time-dependent) extra-axonal space contribution to the total signal, consistent with ref.~\citep{RN2}. 
However, $\delta$ was fixed to a single value while $\Delta$ varied, even though $S_{\rm in}$ mostly depends on $\delta$, and $\Delta$-dependence drops out in the Neuman's limit, cf. {\it Appendix A}. 
Omission of the equally important $K_{\rm ex}$ contribution (as well as, possibly, higher-order cumulant terms) has introduced an unknown bias into parameter estimation. 

Here, we limited our analysis to $b\leq 0.5\,\mathrm{ms/\mu m^2}$ to stay in the linear, DTI regime of \eqref{eq:dmri-signal}.
We therefore cannot estimate $A$ and compartment fractions separately; such estimation would require a systematic measurement of both $\Delta$ and $\delta$ dependencies at higher $b$, and including higher-order cumulants into the model for $S_{\rm ex}(\Delta,\delta;b)$. This is beyond the scope of the present work.   
We also attempted to fit to scan 1 data a hybrid model $D = f_{\rm in}D_{\rm in}(\Delta,\delta) + f_{\rm ex}D_{\rm ex}(\Delta,\delta)$, including finite axonal radius histogram in the van Gelderen's framework of $D_{\rm in}(\Delta,\delta)$; fitting results were unstable, and corresponding parameters were highly dependent on their initial values, signifying a ``shallow direction'' in the fitting landscape. Such spurious parameter correlation should be expected from similar functional forms of the extra-axonal model and of the intra-axonal (van Gelderen) model for large inner radii, cf. Fig.~\ref{fig:vgel-wm} for large $\eta$. 

To simplify models and interpretations, we ignored the fiber orientation dispersion, which is generally non-negligible in the brain WM \citep{Zhang2011-ADM,RN4,Veraart2016-Univ}. 
In the future, it may be possible to consistently factor out this dispersion by generalizing the rotationally-invariant parameter estimation \citep{rotinv,baydiff} onto  time-dependent diffusion propagators.

\subsection{Correlation of dMRI with axonal conduction velocity}


Generally, thicker axons have higher axonal conduction velocity (ACV). 
Within the neuroscience community, it is still an open question whether it is inner or outer axonal diameter, or some combination of both, that determine ACV most definitively.  
Hursh \citep{RN20} observed that, in the peripheral nerve of cats and kittens, the ACV was linearly correlated with the outer axonal diameter. Rushton, and Waxman \& Bennett \citep{RN19,RN21} reanalyzed Hursh's data, and all concluded that ACV is proportional to the outer diameter. However, Sanders and Whitteridge's results in rabbit's peroneal nerve showed that the myelin sheath thickness, i.e. the difference between outer and inner radii, had the highest correlation with ACV, rather than inner and outer diameters separately \citep{RN22}. Arbuthnott \etal \citep{RN23} studied the peripheral nerve of cat and suggested that conduction velocity is proportional to inner diameter; however, they did not measure the conduction velocity in this study, and the conclusion was made based on their theoretical discussion. To estimate the ACV in the human brain, Aboitiz \etal assumed that inner diameter has a linear relationship with ACV; the proportionality constant is 8.7 mm/ms per $\mu$m of inner diameter, which is calculated in the peripheral nervous system \citep{RN9,RN24}. 

The advent of in vivo dMRI has offered an exciting proposition to map axonal diameters, 
and to study in vivo the decades-old relation between axonal sizes and ACV. 
In 2014, based on the AxCaliber interpretation of dMRI, Horowitz \etal  \citep{RN25} estimated apparent inner axonal diameters in the in vivo human brain, and displayed their correlation with ACV measured with electroencephalography;  the estimated proportionality constant was close to the value used by Aboitiz \etal \citep{RN9,RN24} The finding was subsequently criticized by Innocenti, Caminiti and Aboitiz \citep{RN26} since the estimated inner diameter was much larger than histological observations, and the measured interhemispheric transfer time was much shorter than the value in previous literature.
This debate presents an interesting scientific question: Can one rationalize fairly strong apparent correlations between dMRI and ACV observed by Horowitz \etal  \citep{RN25} with the inconsistencies of inner diameter estimation methodology?

The relevance of the nontrivial dMRI signal from the extra-axonal space leads us to posit that the correlation uncovered by Horowitz \etal \citep{RN25} is, to the leading order ${\cal O}(b)$, between the strength of time dependence ($c$ or $c'$ in \eqref{eq:D-neu} or \eqref{eq:D-ex}), and ACV.  Interpreting the strength of time dependence as inner diameter or extra-axonal packing correlation length depends on the model selection.
Our present model selection results suggest re-interpreting dMRI axonal diameter mapping in terms of the dominant extra-axonal contribution, defined in terms of the ``disorder strength'' $A$, and the related axonal packing correlation length 
$l_c^\perp \sim\sqrt{A} \sim \sqrt{c'}$ estimating  {\it outer} diameters.   Selecting the extra-axonal model based on our current data is then consistent with the above mentioned correlations\citep{RN20,RN19,RN21,RN22} between, predominantly, the outer axonal diameters and ACV.

\section{Conclusions}

We considered the functional form of $D(\Delta,\delta)$ for two plausible biophysical models with mutually exclusive physical assumptions.
We experimentally showed in the in vivo human brain, that the extra-axonal model provides a far better agreement with the measurement, both in terms of the quality of its parameter-free prediction of the measurement with varying $\delta$, and in terms of the qualitative $\ln (1/\delta)$, rather than $1/\delta$, functional form. Varying $\delta$ has revealed a nontrivial low-pass filter effect of the gradient duration on the genuine molecular diffusion coefficient $D(t)$.

Extra-axonal model provides reasonable values of the packing correlation length, which is compatible to the scale of outer axonal diameter. In contrast, intra-axonal model alone overestimates  the inner axonal diameters by at least twofold as compared with histology, which cannot be explained by any reasonable degree of the tissue shrinkage in fixation.

The sensitivity of time-dependent diffusion to packing geometry of the extra-axonal space may serve as a marker for demyelination or axonal loss in neurodegenerative diseases. Our results are also consistent with the correlations between outer axonal diameter and axonal conduction velocity.

\newpage

\section*{Acknowlegements} 
We thank Thorsten Feiweier for developing advanced diffusion WIP sequence and Jelle Veraart for assistance in processing. Research was supported by the National Institute of Neurological Disorders and Stroke of the NIH under award number R01NS088040.



\newpage

{

\appendix
\section*{Appendix A. Intra-axonal model}
\renewcommand{\theequation}{A.\arabic{equation}}

Here we obtain qualitative estimates for signal attenuation within an impermeable cylinder in the GPA, outline exact relations for $D_{\rm in}(\Delta,\delta)$ in different limits, and estimate when GPA breaks down. 



\subsection*{Mapping onto transverse relaxation}
Fundamentally, dMRI is a measurement of transverse NMR relaxation in the applied diffusion gradient. 
Each spin, following its Brownian path ${\bf x}(\tau)$, contributes the precession phase 
$e^{-i\phi(t)}$, $\phi(t) = \int_0^t \! \Omega\big({\bf x}(\tau), \tau\big) \, {\rm d}\tau$, 
where $\Omega({\bf x},\tau)$ is the local Larmor frequency offset (relative to $\gamma B_0$), that also depends on time $\tau$ explicitly due to the time-varying applied gradient. The dMRI signal $S = \langle e^{-i\phi}\rangle \equiv p(\lambda)|_{\lambda=1}$, given by the average over all spins in a voxel, is, effectively, the Fourier transform  
$p(\lambda) = \langle e^{-i\lambda \phi} \rangle$ of the probability density function ${\cal P}(\phi)$ of all possible precession phases $\phi(t)$, 
where $\langle \dots \rangle$ is the average with respect to ${\cal P}(\phi)$. 

\subsection*{Wide-pulse limit in the GPA}
Generally, the form of ${\cal P}(\phi)$ is quite complicated, and is mediated by the diffusion \citep{kiselev1998,jensen-chandra-2000,sukstanskii2003,sukstanskii2004,emt-jmr}. 
Fortunately, in the wide-pulse limit $\delta \gg t_c$, the problem of finding its Fourier transform $p(\lambda)$ simplifies, as the problem maps onto that of transverse relaxation in the diffusion-narrowing regime (equivalent to the GPA). In this limit, the time $t_c$ to diffuse across an axon of radius $r$ provides the correlation time, beyond which the contribution to the precession phase $\phi$ for each spin gets randomized.  
It is then natural to split each Brownian path ${\bf x}(\tau)$ into $N = t/t_c \gg 1$ steps of duration $t_c$, such that the total phase can be estimated as
$\phi \sim \sum_{n=1}^N \phi_n$, where each $\phi_n \sim \Omega \cdot t_c$ can be treated as an {\it independent} random variable with zero mean and variance $\langle \phi_n^2 \rangle \sim (\Omega\cdot t_c)^2$; here $\Omega \sim g\cdot r$ is a typical value of the Larmor frequency inhomogeneity across an axon imposed by the applied gradient $g$. When the number $N$ of independent ``steps'' becomes large, the Central limit theorem (CLT) tells that the characteristic function 
$p(\lambda) \simeq e^{-i\lambda\langle\phi\rangle-\lambda^2\langle \phi^2\rangle_c/2}$ 
approaches that of the Gaussian distribution, with the higher-order cumulants being less relevant. 
Moreover, according to the CLT, the mean values and variances from the independent steps add up, i.e. 
$\langle \phi \rangle \equiv 0$, and $\langle \phi^2\rangle_c \equiv \langle \phi^2\rangle - \langle \phi\rangle^2 \sim N \langle \phi_n^2\rangle \sim \Omega^2 \, t_c \cdot t$, 
such that $S_{\rm in}\sim e^{-R_2^*\cdot t}$, with effective $R_2^* \sim \Omega^2 \, t_c$, 
cf. refs.~\citep{kiselev1998,jensen-chandra-2000,sukstanskii2003,sukstanskii2004,emt-jmr}. 
In our case, it is the total pulse duration $t = 2\delta$ that matters; note that the inter-pulse duration $\Delta\geq\delta$ does not enter these considerations, as long as $\delta \gg t_c$, since $\Omega({\bf x},\tau)\equiv 0$ and no transverse relaxation occurs during the time when the gradient is off.  
Hence, the ${\cal O}(g^2)$ attenuation inside an axon scales as 
$-\ln S_{\rm in} \simeq \frac12 \langle \phi^2\rangle_c \sim (g^2 r^4/D_0) \cdot \delta$, 
which indeed agrees with the 1974 exact calculation of Neuman \citep{RN7}
\begin{equation} \label{eq:D-in-neu}
-\ln S_{\rm in} = \frac{7}{48} \cdot \frac{g^2 r^4}{D_0} \cdot \delta + {\cal O}(g^4) \,, 
\end{equation}
where the coefficient $7/48$ is specific to the assumed perfectly circular cylinder cross-section. 

Factoring out $b$ in \eqref{eq:D-in-neu}, cf. \eqref{eq:dmri-signal}, leads to the intra-axonal contribution in \eqref{eq:D-neu}. 
The corresponding $D_{\rm in}$ is about $2-5 \times 10^{-4}\, \mu$m$^2$/ms 
for  $r\sim1\,\mu$m, $D_0=1.5\,\mu$m$^2$/ms, and $\delta = 20\,$ms, being 
much smaller than the measured diffusivity change in our experiment;  
to account for the observed diffusivity variation over diffusion times, apparent radii $\bar r$ need to be much larger, cf. Table~\ref{tab:ex-neu}.

The estimated shrinkage factor for apparent radii ${\bar r}$ in the Neuman's regime is 
\begin{equation} \label{eq:shrink-factor}
\bar{\eta} \equiv \frac{\bar r}{1.48 \, \mu {\rm m}}\,,
\end{equation}
where the denominator is the apparent radius $\left( \langle r^6 \rangle / \langle r^2 \rangle \right)^{\frac{1}{4}}$ calculated via the histology histogram \citep{RN10}, the blue area in Fig.~\ref{fig:histogram}.

\subsection*{General solution in the GPA}

When Neuman's assumption $\delta\gg t_{\rm c}$ is not satisfied, one needs to use the general ${\cal O}(g^2)$ solution for signal attenuation inside a cylinder of radius $r$ by van Gelderen \etal \citep{RN6}: 
\begin{multline} \label{eq:S-in-vgel}
\!\!\!\!\!\!\!\!
-\ln S_{\rm in}^{\rm vG}(\Delta,\delta;r) = \frac{2 g^2 r^4}{D_0}  \sum_{m=1}^\infty \frac{ t_c }{\alpha_m^6(\alpha_m^2-1)} \cdot
\left[ 
2\alpha_m^2{\delta \over t_c} - 2 \right. \\ \left.
\!\!\!\!
+ 2e^{-\alpha_m^2\delta/t_c} + 2e^{-\alpha_m^2\Delta/t_c}
- e^{-\alpha_m^2(\Delta - \delta)/t_c} - e^{-\alpha_m^2(\Delta + \delta)/t_c}
  \right]\!\!
\end{multline}
where $\alpha_m$ is the $m^\mathrm{th}$ root of $\mathrm{d}J_1(\alpha)/\mathrm{d}\alpha=0$, 
and $J_1(\alpha)$ is the Bessel function of the first kind; note that $t_c = t_c(r) = r^2/D_0$. 
In the $\delta\gg t_c$ limit, the $\Delta$-dependence drops out, and \eqref{eq:S-in-vgel} approaches \eqref{eq:D-in-neu}. 
In the opposite, narrow-pulse limit $\delta\ll t_c$, $\ln S_{\rm in}(t,\delta;r)|_{\delta = 0} = -bD(t)$, with $D(t)=r^2/(4t)$. 

In our analysis, we incorporate the axonal radius histogram from the corpus callosum of three post-mortem human brains, by Caminiti \etal  \citep{RN10}, Fig.~\ref{fig:histogram}, and allow for the uniform axonal stretching, $r_i \to \eta r_i$, such that the overall intra-axonal signal for a given shrinkage factor $\eta$ is the volume-averaged \eqref{eq:S-in-vgel} 
\begin{equation} \label{eq:Si-in-vgel}
S_{\rm in}(\Delta,\delta;\eta) = \sum_i f_i \, S_{\rm in}^{\rm vG}(\Delta,\delta;\eta r_i)
\end{equation}
with the normalized weights $f_i = h_i r_i^2/\sum_j h_j r_j^2$ given in terms of the histogram bin values $h_i$. 
The effective $D_{\rm in}^{\rm vGel}(\Delta,\delta;\eta)$ is obtained by factoring out the $b$-value from $\ln S_{\rm in}$, cf. \eqref{eq:dmri-signal}. 
The average radius is $\langle r \rangle \sim 0.67\,\mu$m $\times\,\eta$ according to the weights $h_i$. 
The value of $g^2$ is estimated by the $b$-value defined in \eqref{eq:dmri-signal}. 
The intra-axonal model based on van Gelderen \etal's solution then yields 
\begin{equation} \label{eq:D-vgel}
D^{\rm vGel}(\Delta,\delta;\eta) = D_\infty + f_{\rm in} \, D_{\rm in}^{\rm vGel}(\Delta,\delta;\eta)\,.
\end{equation}
This model includes four parameters: $D_\infty$, $f_{\rm in}$, $\eta$, and $D_0$. To stabilize our fitting, we fixed $D_0$ by the value of the axial diffusivity $D_\parallel$ from the diffusion tensor, and fixed $\eta$ at a few values $[1, 1.5, 2, 2.4]$. 
After that, we only have two fitted parameters, $D_\infty$ and $f_{\rm in}$, estimated from scan 1 data, in Table~\ref{tab:vgel}. 
Using these parameters, we predicted the $\delta$-dependence in scan 2 results based on \eqref{eq:D-vgel} without tunable parameters, shown in Fig.~\ref{fig:vgel-wm}.

\subsection*{Beyond GPA}

Unfortunately, there are no exact results for the ${\cal O}(g^4)$ terms and beyond in \eqref{eq:D-in-neu}. Let us estimate this next-order term using similar qualitative considerations as above, and establish where the GPA breaks down. For that, we need to estimate the 4th-order cumulant 
$\langle \phi^4 \rangle_c \equiv \langle \phi^4 \rangle -3\langle \phi^2 \rangle^2$ of the precession phase in the cumulant expansion \citep{Kiselev2010diffusion} of $p(\lambda)$ taken at $\lambda\equiv 1$: 
\begin{equation} \label{cumexp}
\ln S = -\frac1{2!} \langle \phi^2\rangle_c + \frac1{4!} \langle \phi^4\rangle_c - \dots \,.
\end{equation}
By definition of the kurtosis $K$ of the phase distribution ${\cal P}(\phi)$, $\langle \phi^4 \rangle_c = K \cdot \langle \phi^2 \rangle_c^2$.
In the large-$N$ limit, kurtosis scales as $K \sim - 1/N \sim - t_{\rm c}/\delta$ and is {\it negative} as a result of the confined intra-axonal geometry 
(see the derivation in {\it Supplementary Information}, {\it Section I}). 
As a result, we obtain 
\begin{equation} \label{eq:gpa-in-phi}
\langle \phi^4 \rangle_c \sim - \frac{g^4 r^{10}}{D_0^3} \cdot \delta \,. 
\end{equation}
GPA breaks down when $\langle \phi^4\rangle_c  \sim \langle \phi^2\rangle_c$ in \eqref{cumexp}, 
equivalent to $\langle \phi^2\rangle_c \sim \delta/t_c$, from which the breakdown condition, \eqref{eq:gpa-in-condition} in the main text, follows.
For such strong gradients, all terms in cumulant expansion are of the same order, which 
requires development of non-perturbative approaches.

}



\section*{References}
\bibliography{revtex-sample}
\bibliographystyle{elsarticle-harv}

\vspace{13cm}
\newpage
\section*{Supplementary Information}

Section I provides the derivation of the kurtosis in the {\it Appendix A, beyond GPA}. Section II provides supplementary data of five subjects.

\subsection*{I. Kurtosis of the diffusion on a simple lattice}

Considering a molecule randomly walking on a one-dimen\-sional lattice, we assume that, in each step, the molecule has equal probability to walk to the left and the right. Starting from the origin, after walking $N$ steps, the molecule is away from the origin by $n$ steps. The molecule walks $(N+n)/2$ steps to the right and $(N-n)/2$ steps to the left. Therefore, the diffusion propagator is given by
\begin{equation} \notag
G_{n,N}=\frac{1}{2}\cdot \left(\frac{1}{2}\right)^N\cdot \frac{N!}{\left(\frac{N-n}{2}\right)! \left(\frac{N+n}{2}\right)!}\,,
\end{equation}
where the first $1/2$ is a normalization constant such that \\$\sum_{n=-N}^N G_{n,N} = 1$ for $N\gg 1$. Using Stirling's formula for factorials, $n! \approx \sqrt{2\pi n} \left(\frac{n}{e}\right)^n\cdot\left(1+\frac{1}{12N}\right)$, the propagator is approximated by
\begin{equation} \notag
G_{n,N}\approx \frac{1}{\sqrt{2\pi N}} e^{-\frac{n^2}{2N} - \frac{n^4}{12N^3}}\,,
\end{equation}
which is very similar to the propagator of free diffusion except the correction term $\exp\left(-{n^4}/{12N^3}\right)$. Keeping the lowest order terms of the correction term, we approximate
\begin{equation} \notag
G_{n,N}\approx \frac{1}{\sqrt{2\pi N}} e^{-\frac{n^2}{2N}}\left(1-\frac{n^4}{12N^3}\right)\cdot C\,,
\end{equation}
where $C=\left(1-\frac{1}{4N}\right)^{-1}$ is a normalization constant such that $\int_{-\infty}^{\infty} G_{n,N} dn = 1$. Using the above propagator $G_{n,N}$ to calculate $\langle n^4 \rangle$ and $\langle n^2 \rangle$, we obtain the kurtosis
\begin{equation} \notag
K\equiv \frac{\langle n^4 \rangle}{\langle n^2 \rangle^2}-3 \approx -\frac{2}{N} + {\cal O}\left( \frac{1}{N^2} \right)\,.
\end{equation}

\subsection*{II. Supplementary data}

Fig.~\ref{figS:subj-DEL} shows scan 1 result in WM ROIs of five subjects. In ACR, SCR, PCR, PLIC, and splenium of the corpus callosum, $D_\perp$ decreases with $\Delta$, manifesting expected $\Delta$-dependence; in contrast, based on the scan 2 result in Fig.~\ref{figS:subj-del}, the $\delta$-dependence of $D_\perp$ is too subtle to be individually observed in all WM ROIs.

To evaluate the variability between subjects, probability density functions (PDFs) of radial diffusivities of five subjects in WM ROIs are shown in Figs.~\ref{figS:pdf-DEL} and~\ref{figS:pdf-del}. PDFs of five subjects generally overlap in all WM ROIs, indicating that the variability between subjects is small.


\renewcommand\thefigure{S.\arabic{figure}}
\setcounter{figure}{0}

\begin{figure*}
\centering
\includegraphics[width=.8\linewidth]{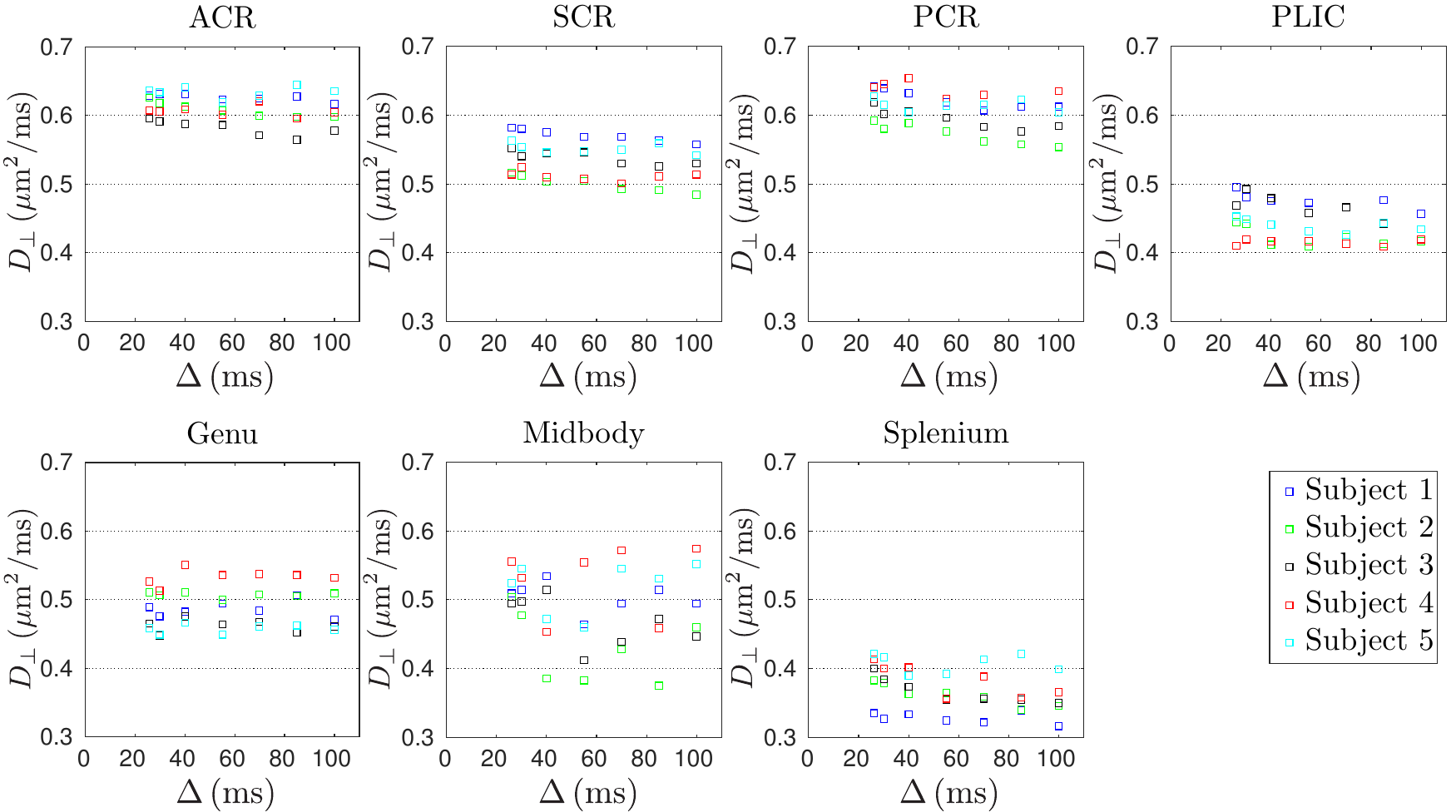}
\caption{Five subjects' radial diffusivities $D_\perp$ in scan 1 within seven WM ROIs with respect to diffusion time $\Delta$.}
\label{figS:subj-DEL}
\end{figure*}

\begin{figure*}
\centering
\includegraphics[width=.8\linewidth]{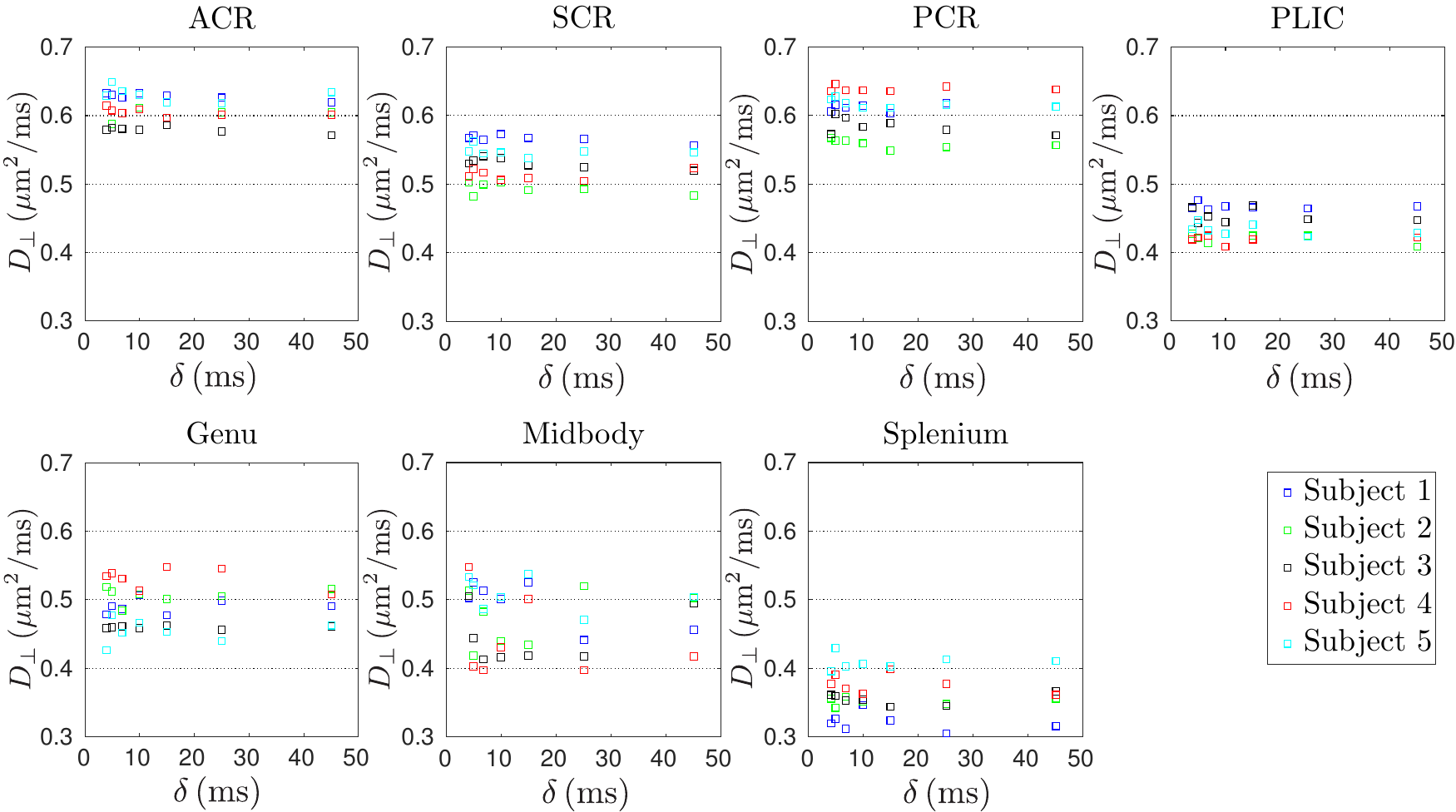}
\caption{Five subjects' radial diffusivities $D_\perp$ in scan 2 within seven WM ROIs with respect to diffusion gradient pulse width $\delta$.}
\label{figS:subj-del}
\end{figure*}

\begin{figure*}
\centering
\includegraphics[width=.8\linewidth]{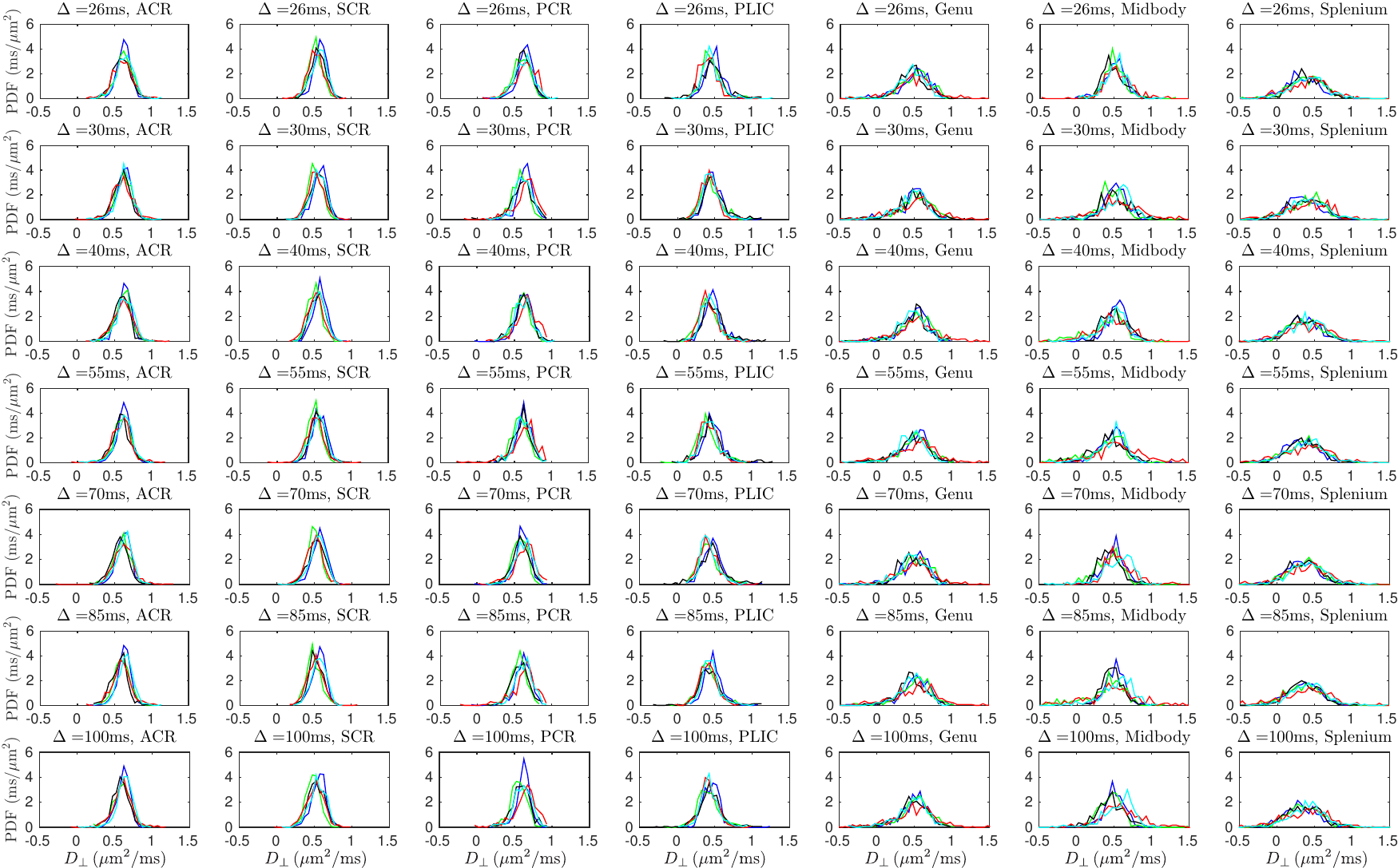}
\caption{Five subjects' probability density functions (PDFs) of radial diffusivities $D_\perp$ in scan 1 within seven WM ROIs with respect to diffusion time $\Delta$.}
\label{figS:pdf-DEL}
\end{figure*}

\begin{figure*}
\centering
\includegraphics[width=.8\linewidth]{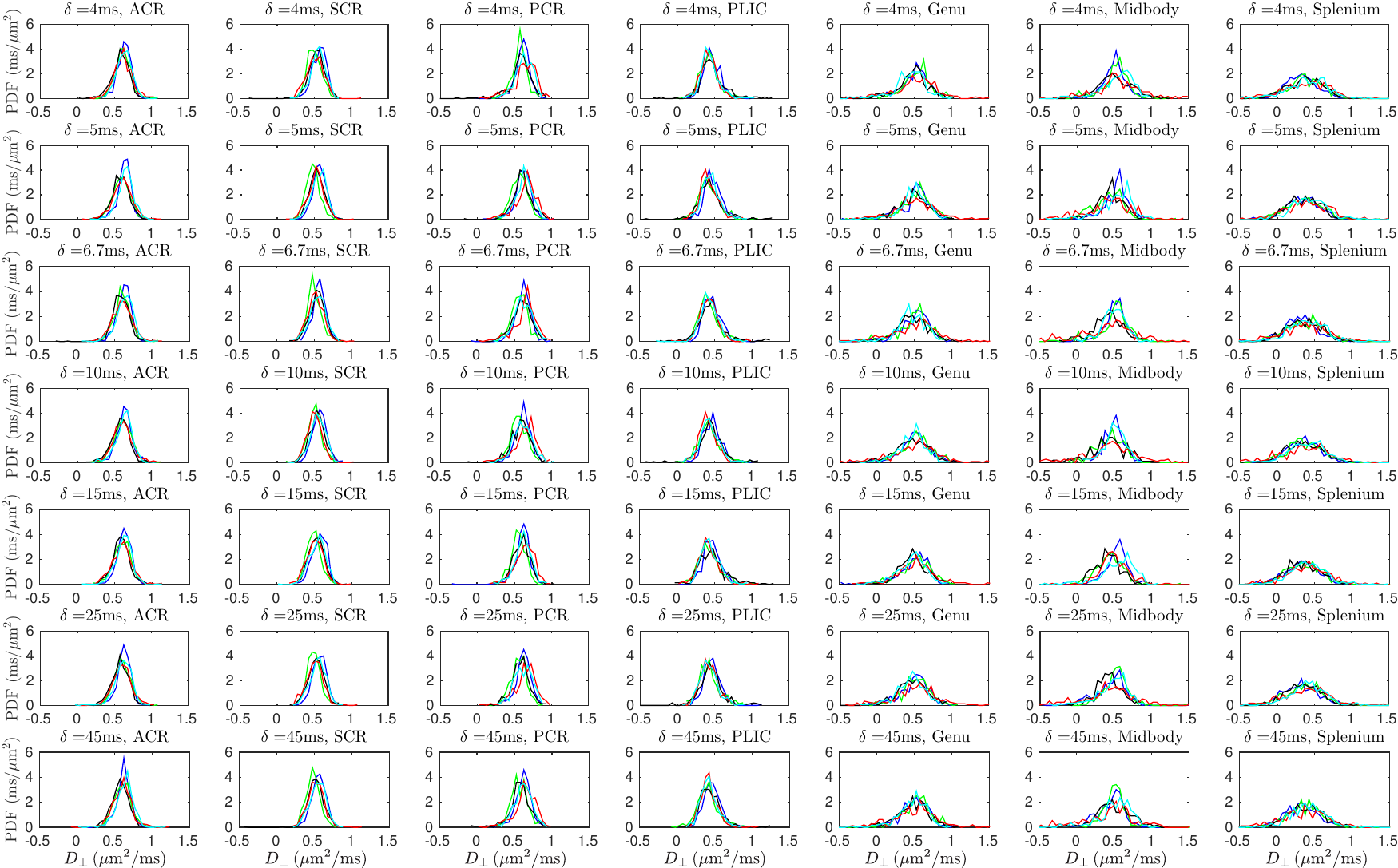}
\caption{Five subjects' probability density functions (PDFs) of radial diffusivities $D_\perp$ in scan 2 within seven WM ROIs with respect to diffusion gradient pulse width $\delta$.}
\label{figS:pdf-del}
\end{figure*}

\end{document}